\documentclass[12pt]{article}
\usepackage{graphicx}
\usepackage{amsmath}
\usepackage{graphics}
\usepackage{epsfig}
\usepackage{amssymb}
\usepackage{color}
 \usepackage{setspace}
\usepackage{extarrows}

\usepackage[authoryear]{natbib}
\usepackage{subfig}

\newcommand{\be}{\begin{eqnarray}}
\newcommand{\ee}{\end{eqnarray}}
\newcommand{\bea}{\begin{eqnarray*}}
\newcommand{\eea}{\end{eqnarray*}}

\newcommand{\phimav}{\Phi_{\text{mav}}}
\newcommand{\mse}{\text{MSE}}

\newtheorem{theorem}{Theorem}[section]

\newtheorem{example}{Example}[section]

\newtheorem{remark}{Remark}[section]

\begin{document}
\title{Optimal designs for frequentist model averaging}

\author{
{\small Kira Alhorn} \\
{\small Technische Universit\"at Dortmund} \\
{\small Fakult\"at Statistik} \\
{\small 44221 Dortmund, Germany } \\
{\small e-mail: kira.alhorn@tu-dortmund.de}\\
\and
{\small Kirsten Schorning} \\
{\small Fakult\"at f\"ur Mathematik} \\
{\small Ruhr-Universit\"at Bochum} \\
{\small 44799 Bochum, Germany} \\
{\small e-mail:kirsten.schorning@ruhr-uni-bochum.de}\\
\and
{\small Holger Dette} \\
{\small Fakult\"at f\"ur Mathematik} \\
{\small Ruhr-Universit\"at Bochum} \\
{\small 44799 Bochum, Germany} \\
{\small e-mail:holger.dette@ruhr-uni-bochum.de}\\
}

\maketitle
\begin{abstract}
We consider the problem of designing experiments for the estimation of a target in regression analysis if there is uncertainty
about the parametric form of the regression function. A new optimality criterion is proposed, which minimizes the asymptotic mean 
squared error  of  the frequentist model averaging estimate by the choice of an experimental design. Necessary conditions for the optimal solution of a locally and Bayesian optimal design 
problem are established. The results are illustrated in several examples  and it is demonstrated that Bayesian optimal designs  
can yield a reduction of the mean squared error  of the model averaging estimator up to $45\%$. 
\end{abstract}
Keywords: Model selection, model averaging, local misspecification, model uncertainty, optimal design, Bayesian optimal deigns
\section{Introduction}
\label{sec1}
\def\theequation{1.\arabic{equation}}
\setcounter{equation}{0}

It is well known that a carefully designed experiment can improve the statistical inference in regression analysis substantially.
Optimal design of experiments is the more efficient the more knowledge about the underlying regression model
is available and an impressive theory has been developed to construct optimal designs under the assumption of a ``given''  regression
model [see, for example, the monographs of  \cite{pukelsheim_optimal_2006}, \cite{atkdonran2007}  and \cite{fedleo2013}].
On the other hand, model selection  is an  important step  in any data  analysis
and these references also  partially discuss the problem of constructing efficient designs to address model uncertainty in the design of experiment.
Because of its importance this problem has  a long history.
Early work dates back to \cite{boxhill1967,stigler1971,atkfed1975a} who determined optimal designs for model discrimination by - roughly  speaking - maximizing  the power of a test  between competing regression models [see also  \cite{ucibog2005,loptomtra2007,wiens2009, dette2009} or \cite{tomlop2010} for some more recent references]. A different line of research in this context  was initiated by
\cite{laeuter1974a}  who proposed a criterion based on a product of the determinants of the information matrices in the various models under consideration,
which  yields efficient  designs  for all  models under consideration. This criterion has been used successfully by \cite{dette1990} to determine efficient designs   for a class of polynomial regression models and by   \cite{biedetpep2006}  to construct efficient designs for binary response models, when there is uncertainty about the form of the link function.
As  these criteria do not reflect model discrimination,
\cite{zentsa2002,atkinson2008b,Tommasi09}
considered   a mixture of  L\"auter-type and discrimination criteria to construct efficient designs for model
discrimination and parameter estimation. An alternative concept to  robust designs with respect to misspecified models consists in the minimization of the 
maximal mean squared error calculated over a class of misspecified models with respect to the design under consideration
[see \cite{wiens_2015} for an overview]. Several authors have worked on this problem and we mention exemplarily 
 \cite{wiens_2008} who derive robust prediction and extrapolation designs 
 or  \cite{konstantinou_2017} who analyze robust designs under local alternatives for survival trials. 
This list of references is by no means complete and there exist many more
papers on this subject. However,   a common feature in most of the literature consists in the fact that either 
(at least implicitly)
the designs are constructed under the assumption that
model selection is  performed by hypotheses testing or
designs are determined with ``good''  properties for a class of competing models.

On the other hand
there exists an enormous amount of literature to perform statistical inference under model
uncertainty, which - to our best knowledge - has not been discussed in the context of optimal experimental design.
One possibility is  to   select  an adequate model from a set of candidate models
and  numerous   model selection criteria have  been developed  for this purpose
[see monographs of  \cite{burand2002}, \cite{konkit2008} and \cite{claeskens_model_2008} among others]. These procedures are widely used and
 have  the advantage  to deliver a single model for the statistical analysis, which make
 them very  attractive for practitoners. However,   there exists a well known post-selection
 problem in this approach  because estimators chosen after model selection  behave usually like mixtures of many potential estimators.
 For example, if $\mu$ is a parameter of interest in a regression model
 (such as a prediction at a particular point, the area under the curve or a specific quantile of the regression model)
 it is known that  selecting a single model and ignoring the uncertainty resulting from the selection process
may  give confidence intervals  for $\mu$ with coverage probability smaller than the nominal value,  see for example
 Chapter 7 in \cite{claeskens_model_2008} for a mathematical treatment
or  \cite{born2015} for a high-level discussion of this phenomenon.

As an alternative several authors  proposed to smooth estimators for the parameter $\mu$   across several models, rather than choosing
a  specific model from the class under consideration and performing the estimation in the selected model.
This approach takes  the additional  estimator variability  caused by model uncertainty adequately into account and
has been discussed intensively in the
Bayesian community, where it   is known as  ``Bayesian model averaging'' [see the tutorial of \cite{hoeting1999} among many others].
 \cite{hjort_frequentist_2003} pointed out several problems with this approach. In particular, they  mentioned the difficulties
 to specify prior probabilities for a class of models  and the  problem  of  mixing together many conflicting prior opinions in the  statistical analysis. As
 an alternative these authors
   proposed  a non-Bayesian approach, which they call   ``frequentist model averaging'' and developed
 asymptotic theory for their method.
There exists   evidence that model averaging improves the
estimation efficiency [see \cite{brei1996} or  \cite{raft:2003}], and recently,
 \cite{schorning_model_2016} demonstrated the superiority of
 model averaging  about  estimation after model selection by an information criterion
 in the context of dose response models.
 These results have recently been confirmed by  \cite{Aoki2017} and \cite{buatois_2018}
 in the context  of  nonlinear mixed effect models.

 The present paper is devoted to the construction of optimal designs if  parameters of interest are  estimated
 under model uncertainty via frequentist model averaging.
 Section \ref{sec2} gives a brief review on model averaging and states the asymptotic properties of this approach under local alternatives.
 The asymptotic properties are used in Section \ref{sec3} to define new optimality criteria, which directly reflect the goal of model averaging.  Roughly speaking, an optimal design for model averaging estimation minimizes the asymptotic mean squared error of the  model averaging estimator under local alternatives.
  In Section  \ref{sec4}  we present a numerical study comparing the optimal designs for model averaging estimation with commonly used designs and demonstrate that the new designs yield substantially  more precise estimates.
Further simulation results which demonstrate that our findings are representative
  can be found in  Section \ref{sec6add}. Finally,  the proofs of the theoretical
   results  are given in Section
  \ref{sec61}.

\section{Model Averaging under local misspecification }
\label{sec2}
\def\theequation{2.\arabic{equation}}
\setcounter{equation}{0}

Model averaging  is  a common technique to estimate  a  parameter of interest, say $\mu$, under model uncertainty.
Roughly speaking this estimate is a weighted average of  the estimates
in the competing models under consideration, where different choices for the weights have
been proposed in the literature [see for example \cite{wass2000} or \cite{hansen_least_2007}
for Bayesian and non-Bayesian model averaging methods]. In this section we briefly
describe  this concept  and  the corresponding asymptotic theory  in the present context, such that the results can be used to construct
optimal designs for model averaging estimation.
 The results  follow more or less  from the statements in \cite{hjort_frequentist_2003} and
 \cite{claeskens_model_2008} and - although we use a  slightly different notation -  any details regarding their derivation are omitted for the sake of brevity.


We assume that $k$ different experimental conditions, say  $x_1, \ldots, x_k$, are chosen in the design space
$\mathcal{X}$, and that  at each  experimental condition  $x_i$  one can observe  $n_i$ responses, say
 $y_{i1}, \ldots, y_{in_i}$ $(i = 1, . . . , k)$.    We  also assume that  for each  $i=1, \ldots, k$  the responses $y_{i1}, \ldots, y_{in_i}$ at experimental  condition $x_i$ are
  realizations of independent  identically (real valued) random variables $Y_{i1}, \ldots, Y_{in_i}$ with
   unknown density $f_{\text{true}}(\cdot |x_i)$.
 Therefore the  total sample size is  given by $n=\sum_{i=1}^{k}n_i $
 and the experimental design problem consists in the choice of $k$ (number of different experimental conditions),
$x_1, \ldots , x_k $ (the experimental conditions) and the  choice  of  $n_1, \ldots , n_k$ (the numbers $n_i$ of
observations taken at each $x_i$), such that the model averaging estimate is most efficient.

To measure  efficiency and to compare different experimental designs we will use  asymptotic arguments
 and  consider the case  $\lim_{n \to \infty} \frac {n_{i}}{n}=\xi_{i} \in (0,1)$ for $i=1, \ldots, k$.
 As common in optimal  design theory  we collect this information  in the matrix
\begin{equation} \label{design}
\xi= \left \{
\begin{array}{ccc}
x_{1} & \dots &x_k \\
\xi_{1} & \dots & \xi_{k}
\end{array}
\right\}.
\end{equation}
Following \cite{claeskens_model_2008}
we assume that $f_{\text{true}}(\cdot|x)$  is contained in a set, say  $\mathcal{S}$, of parametric candidate densities which is constructed as follows.
The first candidate density in $\mathcal{S}$ is given by a  parametric density $f_{\text{wide}}(y|x, \theta, \gamma)$, where the form of $f_{\text{wide}}$
 is assumed to be known,
 $\theta=(\theta_1, \ldots, \theta_p) \in \Theta $ and $\gamma= (\gamma_1, \ldots, \gamma_q) \in \Gamma $
 denote the unknown parameters, which vary in a compact  parameter space, say   $\Theta \times \Gamma \subset \mathbb{R}^p  \times \mathbb{R}^{q} $. The second candidate density is given by the parametric density $f_{\text{narrow}}(y|x, \theta) = f_{\text{wide}}(y|x, \theta, \gamma_0)$,  which is obtained by fixing the parameter value $\gamma$ to a pre-specified (known) value $\gamma_0\in \Gamma$.
 Throughout the paper, we will call $f_{\text{wide}}(y|x, \theta, \gamma)$ the wide density (model) and $f_{\text{narrow}}(y|x, \theta)$ the narrow density (model), respectively.
Additional candidate models  are obtained by choosing certain submodels between the wide density $f_{\text{wide}}(y|x, \theta, \gamma)$ and the narrow density $f_{\text{narrow}}(y|x, \theta)$.
More precisely, for a chosen subset $S \subset \{1, \ldots, q\}$ of indices with cardinality $|S|$, we introduce the projection matrices $\pi_S \in \mathbb{R}^{|S|\times q}$ and  $\pi_{S^c} \in \mathbb{R}^{|S^c|\times q}$ which map a $q$-dimensional vector to its components corresponding to indices in $S$ and $S^c$, respectively.
Using the abbreviations $\gamma_S = \pi_S \gamma$ and $\gamma_{0, S^c} = \pi_{S^c} \gamma_0$, we define the
candidate density $f_S(y|x, \theta, \gamma_S)$ by
\begin{equation}\label{densS}
f_S(y|x, \theta, \gamma_S) = f_{\text{wide}}(y|x, \theta, \gamma_S, \gamma_{0,S^c}).
\end{equation}
Consequently, for the density $f_S(y|x, \theta, \gamma_S)$ the components of $\gamma$  with  indices  in  $S^c = \{1, \ldots, q\}\setminus S$
are fixed to the corresponding components of $\gamma_0$, while the components  with indices in $S$ are considered as unknown parameters.
Note that  $f_{\text{narrow}} =  f_\emptyset$, $f_{\text{wide}}=f_{\{1,\ldots q\}} $ and that in the  most general
case there are $2^q$ possible candidate densities.  As we might not be interested in all possible submodels
we assume that   the competing models are defined by  different sets $S_1,  \ldots,  S_{r}\subset \{1, \ldots, q \}$
(for some  $r\in \{1, \ldots, 2^q\}$). Thus the class  $\mathcal{S}$ of candidate models is given   by
\begin{equation}\label{setS}
\mathcal{S} = \{f_{S_1}(y|x, \theta, \gamma_{S_1}), \ldots,f_{S_{r}}(y|x, \theta, \gamma_{S_{r}})\}~.
\end{equation}
Following \cite{hjort_frequentist_2003}, we consider  local deviations throughout the paper and assume that the ``true''  density is given by
\begin{equation}\label{locmis}
f_{\text{true}, n} (y|x) = f_{\text{wide}}\bigl(y|x,  \theta_0, \gamma_0 + \frac{\delta}{\sqrt{n}}\bigr),
\end{equation}
where the ``true'' parameter values are  given by  $\theta_0 \in \Theta$ and $\gamma_0 + \tfrac{\delta}{\sqrt{n}} \in\Gamma$. Note that the ``true''
density is given by the wide density with a varying value of $\gamma$ which differs from $\gamma_0$ through the perturbation term
$\tfrac{\delta}{\sqrt{n}}$. Thus, for $n$ tending to infinity, it approximates the narrow density $f_{\text{narrow}}(y|x, \theta_0)$.

\begin{example}\label{ex_emax}
{\rm
Consider the case, where  $f_{\text{wide}}(y|x, \theta, \gamma)=f_{S_4}(y|x, \theta, \gamma)$ is a normal density with  variance $\sigma^2$ and
mean function
\begin{equation} \label{emaxmod}
\eta_{S_4}(x, \vartheta, \gamma) = \gamma_1 + \vartheta_1 \frac{x^{\gamma_2}}{x^{\gamma_2} + \vartheta_2^{\gamma_2}},
\end{equation}
where $\theta^T=(\sigma^2, \vartheta_1, \vartheta_2)$, $\gamma^T= (\gamma_1, \gamma_2)$ and the explanatory variable
$x$ varies in an interval, say $[a,b]$.
This model is the well known  sigmoid Emax model and has numerous applications in modelling the dependence of biochemical or pharmacological responses on concentration [see \cite{goutelle_hill_2008} for an overview]. The sigmoid Emax model is especially popular for describing dose-response relationships in drug development [see \cite{ting_dose_2006} among many others].
The parameters in \eqref{emaxmod}  have a concrete interpretation:
 $\gamma_1$  is used to model a  Placebo-effect, $\vartheta_1$ denotes the maximum effect of  $x$  (relative to placebo) and $\vartheta_2$ is the
 value of $x$ which produces half of the maximum effect.
 The so-called Hill parameter $\gamma_2$ characterizes the slope of the mean function $\eta$.
The parameter $\theta$ is included in every candidate model, whereas for the narrow model
the components are fixed as $\gamma_0= (0,1)^T$.
Consequently,
 the narrow candidate model  is a normal density with mean
\begin{equation}\label{emax_narrow}
\eta_{S_1}(x,\vartheta)=  \frac{\vartheta_1 x}{x + \vartheta_2}
\end{equation}
and variance $\sigma^2$.
In this case, $\eta_{S_1}$ is the frequently used Michaelis Menten function, which  is widely utilized to represent   an enzyme kinetics reaction, where enzymes bind
substrates and turn them into products [see, for example, \cite{cornish_fundamentals_2012}].
The two other candidate models between  are obtained by either fixing $\gamma_1=0$ or $\gamma_2=1$
and the corresponding densities are normal densities with mean functions
\begin{equation}\label{emax_between}
\eta_{S_2}(x, \vartheta, (0,\gamma_2) )=  \vartheta_1 \frac{x^{\gamma_2}}{x^{\gamma_2} + \vartheta_2^{\gamma_2}}~,~~\eta_{S_3}(x, \vartheta, (\gamma_1,1) ) =  \gamma_1 + \vartheta_1 \frac{x}{x + \vartheta_2},
\end{equation}
respectively. The latter model is the well known  Emax model which is sometimes also referred to as the hyperbolic Emax model [see \cite{holford_understanding_1981} and  \cite{ting_dose_2006} among others].
Finally, under the local misspecification assumption  \eqref{locmis} the true density $f_{\text{true}, n} (y|x)$ corresponds to a normal distribution with mean
\begin{equation*}
\eta_{\text{true}, n}(x) = \frac{\delta_1}{\sqrt{n}} + \vartheta_1 \frac{x^{1+ {\delta_2}/{\sqrt{n}}}}{x^{1 + {\delta_2}/{\sqrt{n}}} + \vartheta_2^{1 + {\delta_2}/{\sqrt{n}}}}
\end{equation*}
and variance $\sigma^2$.
Typical functionals $\mu$ of interest are the area under the curve (AUC)
\begin{equation} \label{AUCgen}
\mu(\theta, \gamma) = \int_{\mathcal{C}}  \eta (x, \vartheta, \gamma)  dx
\end{equation}
calculated for a given region $\mathcal{C}\subset \mathbb{R}$
or, for a  given  $\alpha \in (0,1)$,
 the ``quantile'' defined by
 \begin{equation} \label{EDPgen}
\mu(\vartheta, \gamma)  = \inf \Big \{ x \in {\cal X }~ \Big |~\frac{ \eta (x, \vartheta, \gamma) - \eta(a, \vartheta, \gamma) }{\eta(b, \vartheta, \gamma) - \eta(a, \vartheta, \gamma)} \geq \alpha   \Big  \} ~.
\end{equation}
The value defined in \eqref{EDPgen} is well-known as $ED_\alpha$, that is, the effective dose at which $100\times \alpha \%$  of the maximum effect
is achieved  [see \cite{ting_dose_2006} or  \cite{bretz_dose_2008}].
}
\end{example}

As pointed out at the end of Example \ref{ex_emax} we are interested in the estimation of a  quantity, say  $\mu(\theta, \gamma)$, where
$\mu: \Theta \times \Gamma \rightarrow \mathbb{R}$
is a differentiable function of  the parameter $(\theta, \gamma)$.  For this purpose we fix  one model $S \in \mathcal{S}$
 in the set of candidate models defined in  \eqref{setS} and  use the estimator $\hat\mu_S = \mu(\hat\theta, \hat \gamma_S, \gamma_{0,S^c})$,
 where $(\hat\theta, \hat \gamma_S)\in \mathbb{R}^{p+|S|}$ is the maximum-likelihood estimator in model $S$.
Under the assumption  \eqref{locmis} of a  local misspecification  and  the common conditions of regularity [see, for example,
\cite{lehmann_theory_1998}, Chapter 6]
it can be shown by adapting the arguments
 in  \cite{hjort_frequentist_2003}  and   \cite{claeskens_model_2008}
 to the current  situation   that   the resulting estimator $\hat{\mu}_S$  satisfies
\begin{equation}\label{asympmlest}
\sqrt{n}\left( \hat\mu_S - \mu(\theta_0,  \gamma_0 + \tfrac{\delta}{\sqrt{n}} )\right)
\xlongrightarrow{\mathcal{D}} \Lambda_S \sim \mathcal{N}(\nu_S(\xi), \tau^2_S(\xi)).
\end{equation}
Here $\xlongrightarrow{\mathcal{D}}$ denotes weak convergence and $ \mathcal{N}(\nu_S(\xi), \tau^2_S(\xi))$ is
 a normal distribution with  variance
\begin{equation}\label{tausquare}
\tau^2_S(\xi)=  \tau^2_S(\xi, \theta_0, \gamma_0) = c^T_SJ^{-1}_S(\xi, \theta_0, \gamma_0)c_S~,
\end{equation}
where $c_S$ is the gradient of $\mu$ with respect to $(\theta, \gamma_S)$, that is,
\begin{equation} \label{cs}
c_S= c_S(\theta_0, \gamma_{0,S}) = \tfrac{\partial}{\partial(\theta, \gamma_S)}\mu(\theta, \gamma_S, \gamma_{0, S^C}) |_{(\theta, \gamma_S)=({\theta_0, \gamma_{0,S}})},
\end{equation}
and $J_S$ the  information matrix $J_S$ in the candidate model $f_S$, that is
\begin{equation}\label{infos}
J_S(\xi, \theta_0, \gamma_{0,S})= \int_{\mathcal{X}} \int
\frac{
\big ( \frac{\partial}{\partial (\theta, \gamma_S)}  f_S(y|x, \theta_0, \gamma_{0,S})\big ) \big (\frac{\partial}{\partial (\theta, \gamma_S)}  f_S(y|x, \theta_0, \gamma_{0,S})\big )^T
}{f_S(y|x, \theta_0, \gamma_{0,S})}
dy \xi(dx).
\end{equation}
The mean $\nu_S(\xi)$ in \eqref{asympmlest} is of the form
\begin{equation*}\label{nus}
\nu_S(\xi)= \nu_S(\xi, \delta, \theta_0, \gamma_0) =c^T L_S(\xi, \theta_0, \gamma_0) \delta~,
\end{equation*}
where
\begin{equation*}\label{cwide}
c= c(\theta_0, \gamma_0) = \tfrac{\partial}{\partial(\theta, \gamma)}\mu(\theta, \gamma) |_{(\theta, \gamma)=(\theta_0, \gamma_0)}
\end{equation*}
is the gradient (with respect to the parameters) in  the wide model,  the matrix $L_S$ is
defined by
\begin{equation}\label{L_S}
L_S(\xi, \theta_0, \gamma_0) =\big (  P_S^T J_S^{-1}(\xi, \theta_0, \gamma_{0,S})  P_S
J(\xi, \theta_0, \gamma_0) - I_{(p+q)\times (p+q)}\big )\begin{pmatrix} 0 _{p\times q} \\ I_{q\times q} \end{pmatrix},
\end{equation}
the matrices   $J_S$ and $P_S$ are given by   \eqref{infos} and
 \begin{equation}\label{Ps}
P_S = \begin{pmatrix} I_{p\times p} & 0 \\ 0 & \pi_S \end{pmatrix},
\end{equation}
respectively, and $J(\xi, \theta_0, \gamma_0)$ denotes the information matrix in the wide model $f_{\text{wide}}$.

 The frequentist model averaging estimator  is now defined  by  assigning  weights
  $g_{S_1}, \ldots, g_{S_r}$, with $\sum_{i=1}^r  g_{S_i} = 1$, to the different candidate models $S_1, \ldots, S_r\in \mathcal{S}$
 and defining
 \begin{equation}\label{mav_est}
\hat\mu_{\text{mav}} = \sum_{i=1}^{r} g_{S_i} \hat\mu_{S_i} ,
\end{equation}
where
  $\hat{\mu}_{S_1}, \ldots, \hat{\mu}_{S_r}$  are the estimators in the different candidate models $S_1, \ldots, S_r\in \mathcal{S}$.
The asymptotic behaviour of the model averaging estimator $\hat{\mu}_{\text{mav}}$ can be derived from   \cite{hjort_frequentist_2003}.
In particular, it can be shown that under assumption \eqref{locmis}  and  the standard regularity conditions a standardized version of $\hat{\mu}_{\text{mav}}$ is asymptotically normally distributed, that is
\begin{equation}\label{asymp_mav_est}
\sqrt{n} \left(\hat\mu_{\text{mav}} - \mu(\theta_0,  \gamma_0 + \tfrac{\delta}{\sqrt{n}} )\right)  \xlongrightarrow{\mathcal{D}} \sum_{i=1}^r
g_{S_i}
\Lambda_{S_i}\sim \mathcal{N}(\nu(\xi, \delta, \theta_0, \gamma_0),  \tau^2(\xi,  \theta_0, \gamma_0))~.
\end{equation}
Here the mean  and variance are given by
\begin{eqnarray}\label{mean_mav}
\nu(\xi, \delta, \theta_0, \gamma_0) &=& \sum_{i=1}^r g_{S_i} \nu_{S_i}(\xi , \delta, \theta_0, \gamma_0) , \\
\label{var_mav}
\tau^2(\xi,  \theta_0, \gamma_0) &=& \sum_{i=1}^r\sum_{j=1}^{r}g_{S_i}g_{S_j} h^T_{S_i}(\xi) J(\xi, \theta_0, \gamma_0) h_{S_j}(\xi) ,
\end{eqnarray}
respectively,   $J(\xi, \theta_0, \gamma_0)$ is the information matrix of the wide model $f_{\text{wide}}$ and the vector $h_S(\xi)$ is given  by
\begin{equation}\label{hs}
h_{S}(\xi) = P_S^T J^{-1}_{S}(\xi, \theta_0, \gamma_{0, S})  c_{S},
\end{equation}
where we used the notation of $c_{S}$, $J_S$ and $P_S$  introduced  \eqref{cs}, \eqref{infos} and  \eqref{Ps}.
The  optimal design criterion for model averaging, which will be proposed in this paper,   is based on an asymptotic
representation of the mean squared error of the estimate $\hat\mu_{\text{mav}} $
derived from \eqref{asymp_mav_est} and will be carefully defined  in the following  section.

\section{An optimality criterion for model averaging  estimation} \label{sec3}
\def\theequation{3.\arabic{equation}}
\setcounter{equation}{0}

Following \cite{kiefer_general_1974} we call the quantity $\xi$ in \eqref{design} an approximate design on the design space $\mathcal{X}$. This means that the support points ${x_{i}}$  define the distinct dose levels  where observations are to be taken and the weights $\xi_{i}$ represent the relative proportion of responses at the corresponding support point $x_{i}$
($i=1,\ldots , k$).
For  an approximate design $\xi$  and  given total sample size   $n$ a rounding procedure is applied to obtain integers $n_{i} $ ($i=1,\ldots,{k})$  from the not necessarily integer valued quantities $\xi_{i}n$  [see, for example \cite{pukelsheim_optimal_2006}, Chapter 12], which define the number of observations at $x_i $  ($i=1, \ldots, k$).

If the observations are taken  according to an approximate design $\xi$  and  an appropriate rounding procedure is used such that $n_i /n  \to \xi_i$  as $n \to \infty$,  the asymptotic mean squared error of the model averaging estimate $\hat\mu_{\text{mav}}$
of the parameter of interest $\mu(\theta_0, \gamma_0 + \delta/\sqrt{n})$
can be obtained from the discussion in Section \ref{sec2}, that is
\begin{equation}\label{asymp_mav_mse}
\Phi_{\text{mav}}(\xi, g, \delta, \theta_0, \gamma_0) =
 \nu^2(\xi, \delta, \theta_0, \gamma_0) + \tau^2(\xi, \theta_0, \gamma_0) \approx n \cdot  \text{MSE} (\hat\mu_{\text{mav}}) ,
\end{equation}
where the variance  $\tau^2(\xi, \theta_0, \gamma_0)$ and the bias $\nu(\xi, \delta, \theta_0, \gamma_0)$ are defined in
equations \eqref{mean_mav} and \eqref{var_mav}, respectively.  Consequently,   a ``good" design for the  model averaging
estimate \eqref{mav_est} should give ``small'' values  of $\Phi_{\text{mav}}$.  Therefore, for a given  finite set $\mathcal{S}$ of candidate
models
$f_S$  of the form \eqref{densS} and weights $g_S$
 a design $\xi^*$ is called {\it  locally  optimal design for model averaging estimation of the parameter $\mu$}, if  it minimizes the function $\Phi_{\text{mav}}(\xi, g, \delta,\theta_0, \gamma_0)$
 in \eqref{asymp_mav_mse}  in the class of  all approximate designs  on $\mathcal{X}$.
Here the term ``locally'' refers to the seminal paper of  \cite{chernoff_locally_1953} on optimal designs for nonlinear regression models.

Locally model averaging optimal designs address uncertainty only  with respect to the model $S$ but require prior information for the parameters
$\theta_0, \gamma_0$ and $\delta$.  While such knowledge might be available in some circumstances [see, for example, \cite{dette2008} or \cite{bretz_practical_2010}]
sophisticated design strategies have been proposed in  the literature, which require less precise knowledge about the model parameters,
 such as sequential, Bayesian or standardized maximin optimality criteria  [see   \cite{pronwalt1985,chaloner1995} and
\cite{dette1997}   among others]. Any of these methodologies can be used to construct efficient
robust designs for model averaging and for the sake of brevity we restrict ourselves to Bayesian optimality criteria.
 \\
Here we address the uncertainty with respect to the unknown model parameters  by a prior distribution, say $\pi$, on
$ \Theta \times \Gamma $   and call a design $\xi^*$ {\it Bayesian optimal for model averaging  estimation of the parameter $\mu$ with respect to the prior $\pi$} if it
minimizes the function
 \begin{equation}\label{mav_mse_bayes}
\Phi^\pi _{\text{mav}}(\xi) = \int_{\Theta \times \Gamma } \Phi_{\text{mav}}(\xi, g, \delta,\theta, \gamma) \pi(d\theta, d\gamma),
\end{equation}
where the function $ \Phi_{\text{mav}}$ is defined in \eqref{asymp_mav_mse} (we assume throughout this paper that the integral exists).

Locally and Bayesian optimal designs for model averaging have to be calculated numerically in all cases of interest, and we present several examples
in Section \ref{sec4}. Next, we state necessary conditions
for $\Phi_{\text{mav}}$- and $\Phi_{\text{mav}}^\pi$- optimality.
The proofs are given in the Section \ref{sec61}.

\begin{theorem}\label{mav_necess_cond}
If the design $\xi^*$ is a  locally  optimal design for
model averaging estimation of the parameter $\mu$, then
the inequality
\begin{equation} \label{nec_cond}
-2\nu(\xi^*,\delta,\theta_0,\gamma_0) D_1(\xi^*, x,\delta,\theta_0, \gamma_0) - D_2(\xi^*, x,\theta_0, \gamma_0)  \leq 0
\end{equation}
holds for all  $x \in \mathcal{X}$, where $\nu(\xi^*,\delta,\theta_0,\gamma_0)$ is defined by \eqref{mean_mav} and the
functions $D_1$ and $D_2$ are given by
\begin{eqnarray} \label{deriv_nu}
D_1(\xi^*, x,\delta,\theta_0, \gamma_0)  &= &\sum_{j=1}^{r} g_{S_j} c^T P_{S_j}^T
 J_{S_j}^{-1}(\xi^*, \theta_0, \gamma_{0,{S_j}}) \Big (P_{S_j}
 J(\xi_x, \theta_0, \gamma_0)   \\
\nonumber
 & & - J_{S_j}(\xi_x, \theta_0, \gamma_{0,{S_j}}) J_{S_j}^{-1}(\xi^*, \theta_0, \gamma_{0,{S_j}})
 P_{S_j}
  J(\xi^*, \theta_0, \gamma_0) \Big ) \begin{pmatrix} 0_p \\ \delta\end{pmatrix}, \\
   D_2(\xi^*, x,\theta_0, \gamma_0)  &= &\sum_{i,j=1}^{r}g_{S_i} g_{S_j}\left({h}^T_{S_i}(\xi^*) \{J(\xi^*, \theta_0, \gamma_{0})+ J(\xi_x, \theta_0, \gamma_{0})\} {h}_{S_j}(\xi^*) \right. \label{deriv_tau} \\
  \nonumber
& &\left. -2 \tilde{h}^T_{S_i}(\xi^*, \xi_x) J(\xi^*, \theta_0, \gamma_{0}){h}_{S_j}(\xi^*)  \right) ,
\end{eqnarray}
where the vector ${h}_{S}(\xi)$ is defined by \eqref{hs}, the vector $\tilde{h}_{S}(\xi^*, \xi)$ by
\begin{equation*}\label{hstilde}
\tilde{h}_{S}(\xi^*, \xi) =  P^T_S J^{-1}_{S}(\xi^*, \theta_0, \gamma_{0, {S}}) J_{S}(\xi, \theta_0, \gamma_{0, {S}}) J_{S}^{-1}(\xi^*, \theta_0, \gamma_{0, {S}})  c_{S},
\end{equation*}
and the information matrix $J_S(\xi, \theta_0, \gamma_0)$ by \eqref{infos}, respectively. The design $\xi_x$ denotes the Dirac measure at the point $x \in {\cal X}$.\\
Moreover, there is equality in \eqref{nec_cond} for every point $x$ in the support of $\xi^*$. 
\end{theorem}

\begin{theorem}\label{bayes_mav_necess_cond}
If a  design $\xi^*$ is  Bayesian optimal for model averaging  estimation of the parameter $\mu$ with respect to the prior $\pi$, then
\begin{equation} \label{nec_cond_bayes}
d_\pi  (x , \xi^*)  =
 \int_{\Theta \times \Gamma } -2\nu(\xi^*, \delta, \theta, \gamma) D_1(\xi^*, x,\delta,\theta, \gamma)   -   D_2(\xi^*, x,\theta, \gamma) \pi(d\theta, d\gamma)  \leq 0
\end{equation}
holds for all $x \in \mathcal{X}$, where the derivatives $D_1$ and $  D_2 $ are given by \eqref{deriv_nu} and \eqref{deriv_tau}, respectively.\\
Moreover, there is equality in \eqref{nec_cond_bayes} for every point $x$ in the support of $\xi^*$. 
\end{theorem}

 The derived conditions of Theorem \ref{mav_necess_cond} and Theorem \ref{bayes_mav_necess_cond} can be used in the following way: If a numerically calculated design does not satisfy inequality \eqref{nec_cond}, it will not be locally optimal for model averaging estimation of the parameter $\mu$ and the search for the optimal design has to be continued.
 Note  that   the functions $\Phi_{\text{mav}}$
and $\Phi_{\text{mav}}^\pi $  are not convex  and therefore  sufficient conditions for optimality are not available.

\begin{remark} \label{remark21}
{\rm
 Note  that  \cite{hjort_frequentist_2003} also consider model averaging  using  random weights $g_{S_1}(Y_n), \ldots, g_{S_r}(Y_n)$
  depending on the data $Y_n=(Y_{11}, \ldots, Y_{1n_1}, \ldots, Y_{kn_k})$ in the
 definition  of  the  estimator $\hat{\mu}_{\text{mav}}$ in \eqref{mav_est}. Typical examples are smooth AIC-weights
 \begin{equation}\label{smooth_aic_weights}
g^{{\footnotesize \mbox{smAIC}}}_{S_j} (Y_n) = \frac{\exp(\tfrac{1}{2} \mbox{AIC}(S_j|Y_n))}{\sum_{i=1}^{r} \exp(\tfrac{1}{2} \mbox{AIC}(S_i|Y_n))}.
\end{equation}
 which are based on the AIC-scores
 $$
  \mbox{AIC}(S_i|Y_n) = 2\ell_{S_i}(\hat{\theta},\hat{\gamma}_{S_i}) - 2 d_{S_i} ,
  $$
  where $\ell_{S_i}(\hat{\theta},\hat{\gamma}_{S_i})$ denotes the log-likelihood function of model $S_i$ evaluated in the maximum likelihood estimator $(\hat{\theta},\hat{\gamma}_{S_i})$ and $ d_{S_i}$ is the number of parameters to be estimated in model $S_i$ ($i=1, \ldots , r$) [see \cite{claeskens_model_2008},~Chapter 2].
Moreover, the estimator of the target $\mu$  which is based on model selection by AIC
can also be rewritten in terms of a model averaging estimator by using random weights of the form
\begin{equation} \label{aic_weights}
g^{{\footnotesize \mbox{AIC}}}_{S_j}(Y_n)= I\{S_j = S_{{\footnotesize \mbox{AIC}}}\},
\end{equation}
where $I\{A\}$ is the indicator function of the set $A$ and $S_{{\footnotesize \mbox{AIC}}}$ denotes the model with the greatest AIC-score among the candidate models. For further choices of model averaging weights see for example \cite{buckland_model_1997}, \cite{hjort_frequentist_2003} or \cite{hansen_least_2007}.
In general,  the case of random weights in model averaging estimation  is  more difficult
 to handle  and in general the asymptotic distribution is not  normal [see  \cite{claeskens_model_2008}, p.~196].
As a consequence  an explicit calculation of the asymptotic bias and variance  is not available.

 From the design perspective it therefore seems to be reasonable to consider the case of fixed weights,
 for which the asymptotic properties  of model averaging estimation under local misspecification are well understood
 and  determine efficient designs for this estimation technique.
Moreover, we also demonstrate in Section \ref{sec4} and in the appendix (see Section \ref{sec6})
that model averaging estimation with fixed weights often shows a better performance than model averaging with
 smooth AIC-weights and that
  the optimal designs derived under the assumption of fixed weights also
 improve the current state of the art for model averaging using random weights. }

 \end{remark}

\section{Optimal designs for model averaging} \label{sec4}
\def\theequation{4.\arabic{equation}}
\setcounter{equation}{0}

In this section, we  investigate the performance of optimal designs   for   model averaging estimation of a parameter $\mu$.
For this purpose, we consider several examples from the literature, and
compare the  Bayesian optimal designs for model averaging estimation of a parameter $\mu$  with commonly used uniform designs
by means of a simulation study. Thoughout this paper we use  the
 cobyla algorithm for the minimization of the criterion
 $\Phi^{\pi}_{\text{mav}}(\xi)$ defined in \eqref{mav_mse_bayes}  [see \cite{powell_direct_1994} for details].

\subsection{Estimation of the ED$_\alpha$ in the  sigmoid Emax model }\label{sec41}

\begin{figure}[!t]
\centering

\subfloat{\includegraphics[width=0.45\textwidth]{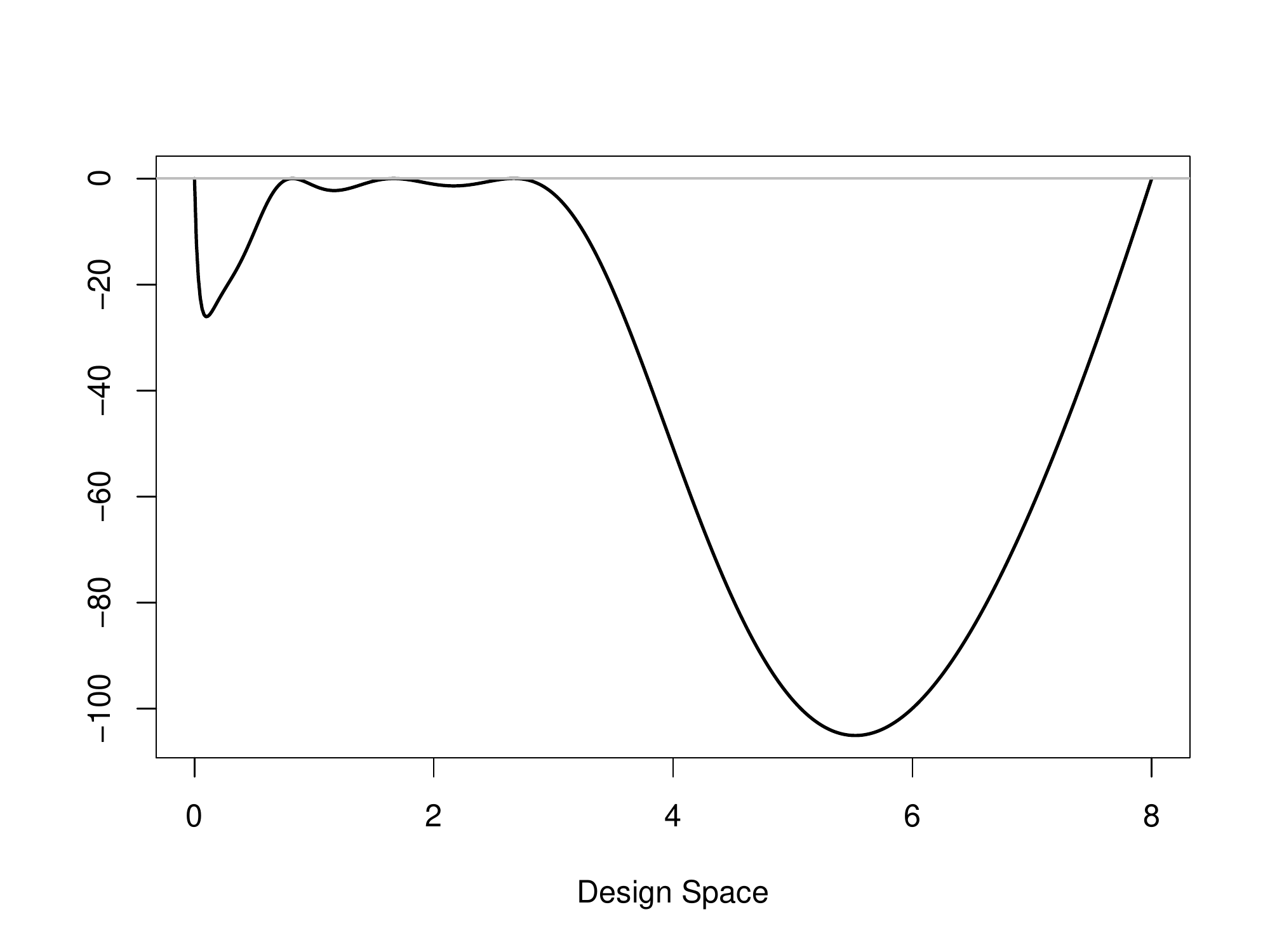}
} \qquad
\subfloat{\includegraphics[width=0.45\textwidth]{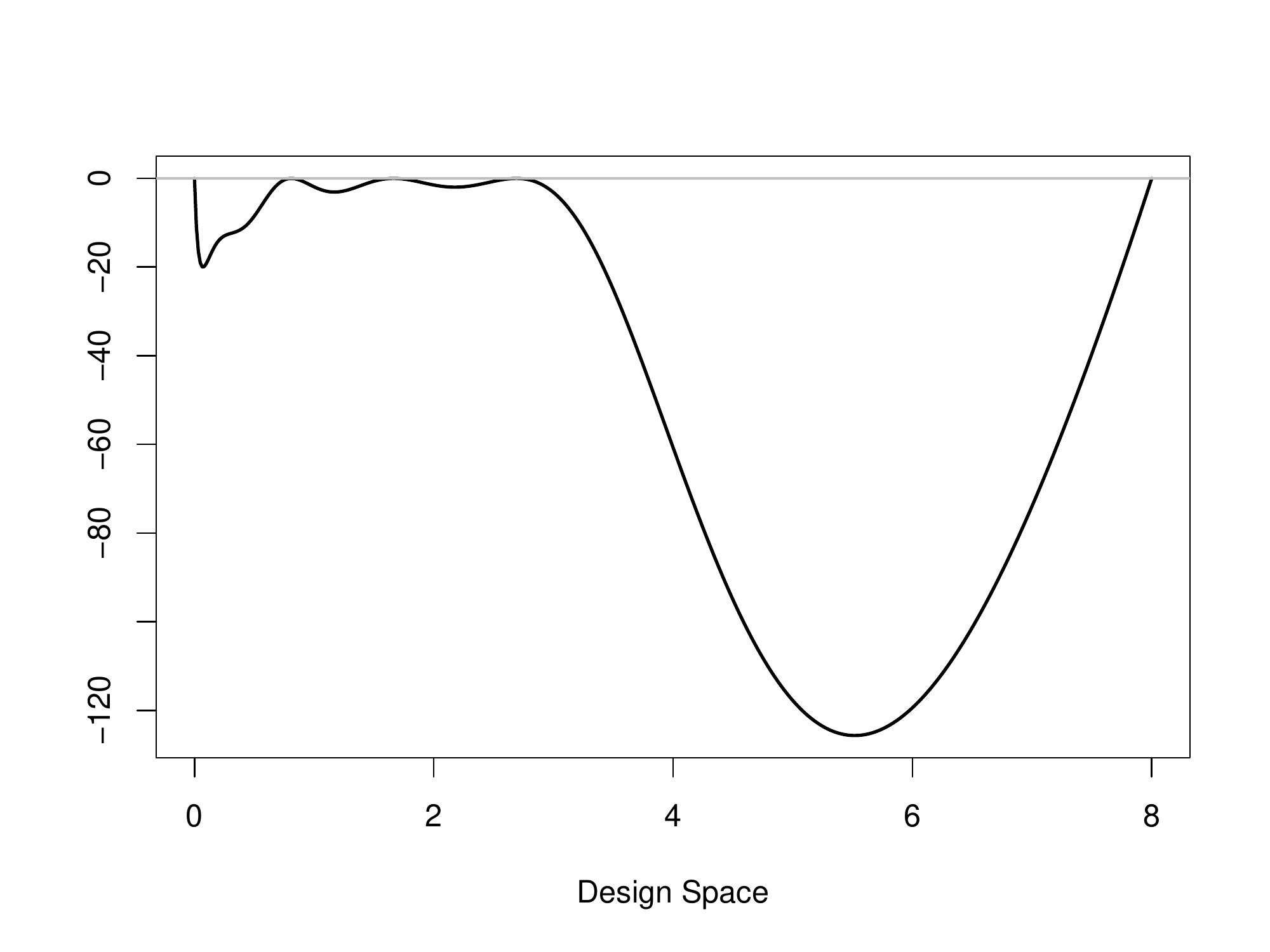}
} \\
\caption{\it The function $d_\pi $  in  \eqref{nec_cond_bayes} evaluated for  the design $\xi^*_A$  in
 \eqref{opt_des_emax} (left panel) and the design  $\xi^*_B$  in
 \eqref{opt_des_emax_2} (right panel).}
\label{fig_deriv_emax_edp}
\end{figure}

We consider the situation introduced in Example \ref{ex_emax}, where
the underlying density is a normal distribution with variance $\sigma^2$ and different regression functions
are under consideration for the mean. More precisely,  the set $\mathcal{S}$ contains $r=4$ candidate models
which are  defined by the different mean functions \eqref{emaxmod}, \eqref{emax_narrow} and \eqref{emax_between}, respectively.
The parameter of interest $\mu$ is the $\text{ED}_{0.6}$ defined in \eqref{EDPgen},
 which is estimated by an appropriate model averaging estimator.
The design space is given by the interval $\mathcal{X}= [0, 8]$ and we assume that $n=150$ observations can be taken in the experiment.
\\
We determine a Bayesian optimal design for model averaging   estimation of the $\mbox{ED}_{0.6}$.
As the Emax model is linear in the
parameters ${\vartheta_1}$ and ${\gamma_1}$,
the optimality criterion does not depend on ${\vartheta_1}$ and ${\gamma_1}$ and no prior information is required for these parameters.
For the parameters $( \vartheta_2 , \gamma_2 ) $  we choose
independent uniform priors
  $\pi_{\vartheta_2}$ and $\pi_{\gamma_2}$  on the sets $\{0.79,1.79,2.79\}$ and $\{1,2,3\}$, respectively,
  and  the variance ${\sigma^2}$ is fixed as $\sigma_0^2 = 4.5$ (note that one can choose a prior for $\sigma^2$ as well).
   Finally, under the local misspecification assumption we set  $\delta$ such that  $\delta/\sqrt{n}=\delta/\sqrt{150}= (0.1, 1)^T$.
\\
We first consider equal  weights for the model averaging estimator, that is   $g_{S_i}=0.25$, $i=1, \ldots, 4$.
The Bayesian optimal design for  model averaging estimation of the $\mbox{ED}_{0.6}$ is given by
\begin{align} \label{opt_des_emax}
\xi^*_A = \left\{
\begin{array}{ccccc}
0 & 0.819 & 1.665 & 2.669 & 8 \\
0.105 & 0.138 & 0.199 & 0.273 & 0.285
\end{array}
\right\},
\end{align}
and satisfies the necessary condition of optimality  in Theorem \ref{bayes_mav_necess_cond}
[see the left panel of Figure \ref{fig_deriv_emax_edp}]. Note that the design  $\xi_A^*$ defined by \eqref{opt_des_emax} would not be optimal if the inequality was not satisfied.

In order to investigate the properties of the different designs for model averaging estimation we have conducted a
simulation study, where we compare the Bayesian optimal design \eqref{opt_des_emax} for model averaging estimation of the $\text{ED}_{0.6}$  with two uniform designs
\begin{eqnarray} \label{com_used_des1}
	\xi_1 &= &
	\left\{ \begin{array}{ccccc} 0 & 2 & 4 & 6 & 8 \\ ^1/_5 & ^1/_5 & ^1/_5 & ^1/_5 & ^1/_5 \end{array}
	 \right\} ~, \\
\label{com_used_des2}
	\xi_2 &=&
	 \left\{ \begin{array}{ccccccccc} 0 & 1 & 2 & 3 & 4 & 5 & 6 & 7 & 8 \\ ^1/_9 & ^1/_9 & ^1/_9 & ^1/_9 & ^1/_9 & ^1/_9 & ^1/_9 & ^1/_9 & ^1/_9 \end{array} \right\},
\end{eqnarray}
which are quite popular in the presence of model uncertainty
[see \cite{schorning_model_2016} and \cite{bornkamp_innovative_2007}].
Note that the design $\xi_1$ is a uniform design with the same number of support points as the optimal design in \eqref{opt_des_emax}, whereas the design $\xi_2$ is a uniform design with more support points.
 Moreover, we also
provide a comparison with two estimators commonly used in practice, namely  the  model averaging estimator based on smooth AIC-weights defined in \eqref{smooth_aic_weights} and  the estimator in the model chosen by AIC model selection, which
is obtained  as a model averaging estimator \eqref{mav_est} using  the weights in  \eqref{aic_weights}. For these estimators we also used observations taken according to
the designs  $\xi_A^*$,  $\xi_1$ and $\xi_2$. As the  approximate designs cannot be implemented directly for $n=150$ observations
a rounding procedure  [see, for example \cite{pukelsheim_optimal_2006}, Chapter 12] is applied to determine the number $n_i$ of observations taken at $x_i$ such that we have in total $\sum_{i=1}^kn_i=150$ observations.  For example, the implemented design obtained from the Bayesian optimal design $\xi_A^*$ for model averaging estimation of the $\text{ED}_{0.6}$ uses
$n_1=16$,  $n_2=21$, $n_3=30$,  $n_4=40$ and $n_5=43$ observations at the points
$0$, $0.819$, $1.165$,  $2.669$ and   $8$, respectively, and implementable versions of the designs $\xi_1$ and $\xi_2$ are obtained similarly.
\\
All results presented  here  are based on $1000$ simulations runs generating in each run $150$ observations
of the form
\begin{equation}\label{sim_values}
y_{ij}^{(l)} = \gamma_1 + \vartheta_1 \frac{x_i^{\gamma_2}}{x_i^{\gamma_2} + \vartheta_2^{\gamma_2}}+ \sigma \varepsilon_{ij}^{(l)}
,i=1,\ldots,k,j=1,\ldots,n_i,
\end{equation}
 for the different designs, where  the   $\varepsilon_{ij}^{(l)}$  are independent standard normal distributed random variables
 and different combinations  of the ``true'' parameter $(\vartheta^T,\gamma^T) = (\vartheta_1,\vartheta_2,\gamma_1,\gamma_2)$
 in \eqref{sim_values} are investigated whereas $\sigma^2=4.5$ is fixed.
In the following discussion we will restrict ourselves to presenting  results for the parameters $(\vartheta_1, \vartheta_2)= (1.81, 0.79)$, $(\gamma_1,\gamma_2)= (0.1, 2)$.
Note that this is the parameter combination  under local misspecification assumption for $\theta_0= (4.5,1.81, 0.79)^T$,
 $\gamma_0=(0, 1)^T$
and $\delta/\sqrt{150}= (0.1, 1)^T$. Further simulation results for other parameter combinations can be found in Section \ref{sec62}.
\\ \noindent
 \begin{table}[t]
 \centering
 \begin{tabular}{| c || c|c|c|}
 \hline
  & \multicolumn{3}{|c|}{estimation method} \\
\cline{2-4}
design  &  fixed weights &  smooth AIC-weights & model selection \\
 \hline\cline{2-4}
$\xi_A^*$  & \textbf{0.355} & \textbf{0.508} & \textbf{0.596} \\
$\xi_1$ & 0.810 & 0.913 & 1.017 \\
$\xi_2$ & 0.637 & 0.846 & 0.994 \\
  \hline
 \end{tabular}
 \caption{\it
 The    mean squared error of the model averaging estimators with weights $g_{S_i}=0.25$, $i=1, \ldots, 4$ (left column), with the smooth AIC-weights  \eqref{smooth_aic_weights} (middle column) and the estimator based on model-selection (right column).
The different rows correspond to different designs. First row: Bayesian optimal design $\xi^*_A $  for model averaging estimation of the $\text{ED}_{0.6}$ defined  in
\eqref{opt_des_emax}. Middle row: uniform design  $\xi_1$  defined in \eqref{com_used_des1}. Third row:  uniform design  $\xi_2$  defined in \eqref{com_used_des2}.}
\label{simulated_mse_values}
 \end{table}
 \noindent 
In each simulation run, the  parameter $\mu= \mbox{ED}_{0.6}$ is estimated  by model averaging using the different designs 
and  the   mean squared error is calculated from all $1000$ simulation runs.
More precisely, if  $\hat\mu^{(l)}_{\text{mav}}$ is the  model averaging estimator for the parameter of interest $\mu= \mbox{ED}_{0.6}$
based on the observations $y^{(l)}_{11}, \ldots, y^{(l)}_{kn_k}$  from  model \eqref{sim_values}  with the  design $\xi$, its   mean squared  error is given by
\begin{equation*} \label{mse_emp}
	\mse (\xi) = \frac{1}{1000} \sum_{l=1}^{1000} \left( \hat{\mu}_{\text{mav}}^{(l)} - \mu_{true} \right)^2,
\end{equation*}
where  $\mu_{true}$ is the  $\mbox{ED}_{0.6}$ in  the
 ``true''   sigmoid Emax model \eqref{sim_values}  with parameters   $(\vartheta^T,\gamma^T) = (1.81, 0.79, 0.1, 2)$.  The simulated mean squared error
of  the model averaging estimator with fixed weights $g_{S_i} = 0.25$, $i=1, \ldots, 4$ for the different designs $\xi^*_A$, $\xi_1$ and $\xi_2$
is shown in the left column of Table \ref{simulated_mse_values}. The middle column of this table shows the   mean squared error of the model averaging
estimator  with the smooth
 AIC-weights in \eqref{smooth_aic_weights}, while the right column gives the corresponding results for the weights in \eqref{aic_weights}, that is
 estimation of  the $ \mbox{ED}_{0.6}$   in the model identified by the AIC for the different designs. The numbers printed in boldface in each column correspond to the smallest mean squared error obtained from the three designs. 
\\
 We observe that  model averaging  always  yields a smaller mean squared error than estimation in the model identified by the AIC.
 For example,  if the design $\xi^*_A$  is used, the mean squared error of the estimator based on model selection is $0.596$, whereas it
 is  $0.355$ and $0.508$  for the  model averaging estimator using  fixed weights and smooth AIC-weights, respectively (see the first row in
 Table \ref{simulated_mse_values}). The situation for the non-optimal uniform designs is similar.
 These results (and also further simulation results presented in Section \ref{sec62}) coincide with the findings of  \cite{schorning_model_2016},
  \cite{Aoki2017} and \cite{buatois_2018}
 and  indicate that  model averaging usually  yields  more precise estimates  of the target than estimators based on model selection.
Moreover, model averaging
 estimation with fixed weights shows a substantially better performance than the model averaging
 estimator with data driven weights. Note that  \cite{wagner_growth_2015} observed a similar effect in the context 
of  principal components augmented regressions.
 \\
  \begin{table}[t]
 \centering
 \begin{tabular}{| c || c|c|c|}
 \hline
 & \multicolumn{3}{|c|}{estimation method } \\
\cline{2-4}
design &  fixed weights &  smooth AIC-weights & model selection  \\
 \hline\cline{2-4}
$\xi_B^*$  & \textbf{0.476} & \textbf{0.502} & \textbf{0.582} \\
$\xi_1$ & 0.915 & {0.900} & {1.014} \\
$\xi_2$ & 0.869 & {0.949} & {1.067} \\
  \hline
 \end{tabular}
 \caption{\it
  The    mean squared error of the model averaging estimators with weights $g_{S_1}=0.1$, $g_{S_2}=0.1$, $g_{S_3}=0.3$ and $g_{S_4}=0.5$ (left column), with the smooth AIC-weights  \eqref{smooth_aic_weights} (middle column) and the estimator based on model-selection (right column).
The different rows correspond to different designs. First row: Bayesian optimal design $\xi^*_B $  for model averaging estimation of the $\text{ED}_{0.6}$ defined  in
\eqref{opt_des_emax_2}. Middle row: uniform design  $\xi_1$  defined in \eqref{com_used_des1}. Third row:  uniform design  $\xi_2$  defined in \eqref{com_used_des2}.}
\label{simulated_mse_values_2}
 \end{table}
 Compared to the uniform designs $\xi_1$ and $\xi_2$ the  optimal design $\xi^*_A$ in \eqref{opt_des_emax}
  yields a reduction of the mean squared error by $56\%$ and $44\%$  for  model averaging estimation with fixed weights.
  Moreover, this design also reduces the  mean squared error of  model averaging estimation with smooth AIC-weights (by  $44\%$ and $40\%$) and
  for estimation in the model identified by the AIC (by $41\%$ and $40\%$).
 \\
As a further example we consider the model averaging estimator   \eqref{mav_est}   of the parameter  $\mbox{ED}_{0.6}$
for the four models in Example \ref{ex_emax} with non-equal weights, that is
    $g_{S_1} = 0.1$ , $g_{S_2} = 0.1$, $g_{S_3} = 0.3$  and  $g_{S_4} = 0.5$. The  Bayesian optimal design for model averaging estimation of the $\text{ED}_{0.6}$
is then given by
\begin{align} \label{opt_des_emax_2}
\xi^*_B= \left\{
\begin{array}{ccccc}
0 & 0.809 & 1.665 & 2.691 & 8 \\
0.152 & 0.120 & 0.175 & 0.279 & 0.274
\end{array}
\right\}.
\end{align}
The necessary condition is depicted in the  right  panel of Figure \ref{fig_deriv_emax_edp}.
 A comparison of the designs $\xi^*_A$ and $\xi^*_B$  in
\eqref{opt_des_emax} and \eqref{opt_des_emax_2} shows   that the support points   are  similar, but that there
 appear  substantial differences in the weights. \\
In the simulation study of this model averaging estimator
 we consider the same parameters as in the previous example.
 The corresponding results can be found in Table  \ref{simulated_mse_values_2} and show a similar but less pronounced picture as for the model averaging estimator
 with uniform weights.
 Model averaging always shows a better performance than estimation in the model selected by the AIC (improvement between $10\%$ and $19\%$ using fixed weights and between $11\%$ and $14\%$ using smooth AIC-weights). Moreover, for the designs $\xi_B^*$ and $\xi_2$
we observe an improvement when using fixed weights instead of smooth AIC-weights for model averaging, but
for the design $\xi_1$ there is in fact no improvement.
 A comparison of the results in  Table  \ref{simulated_mse_values}  and  \ref{simulated_mse_values_2}
shows  that for all designs non-uniform weights  for  model averaging estimation yield a  larger
 mean squared error than uniform weights.
\\
 The Bayesian optimal design $\xi^*_B$ for model averaging estimation of the $\text{ED}_{0.6}$ improves the designs $\xi_1$ and $\xi_2$ by $48\%$ and $45\%$, respectively, if model averaging with fixed (non-uniform weights)
 is used, and by  $43\%-47\%$ for model averaging estimation with smooth AIC-weights and estimation in the model selected by the AIC.
 \\
Simulation results for further parameter combinations in the sigmoid Emax model  can be found in Table \ref{tab_emax_edp_150}  and \ref{tab_emax_edp_150_2} in Section \ref{sec62}.
These results show a very similar picture as described in the previous paragraphs.
We observe that in all considered scenarios model averaging estimation yields a smaller simulated mean squared error than estimation in a model identified by the AIC, independently of the design and parameters under consideration.
 Bayesian optimal designs for model averaging estimation of the $\text{ED}_{0.6}$  yield  a substantially more precise estimation
 than  the uniform designs in almost all cases.
We refer to Section \ref{sec62} for more details.

\subsection{Estimation of the AUC in  the logistic regression model }\label{sec42}

In this section we consider
the logistic  regression model
 \begin{equation} \label{logmod}
\eta_{S_4}(x, \vartheta, \gamma) = \gamma_1 + \frac{\vartheta_1 }{1+ \exp[(\vartheta_2-x)/\gamma_2]},~x \geq 0
\end{equation}
which is frequently  used  in dose-response modeling or modeling population growth
 [see, for example,  \cite{zwietering_modeling_1990}].
 This means we consider a normal distribution with variance $\sigma^2$  and mean (function) given by \eqref{logmod}.
The design space is  given by  $\mathcal{X}=[0, 8]$ and we  are interested in the estimation of  the  area under the curve (AUC) defined in \eqref{AUCgen},
where   the region $\mathcal{C}$ and the design space $\mathcal{X}$ coincide.
In  model   \eqref{logmod} the value
$\eta(0, \vartheta, \gamma) $ is the Placebo-effect, $ \vartheta_1$ denotes the maximum effect (relative to placebo) of the drug
and $\vartheta_2 >0 $ is the dose which produces half of the maximum effect.  The parameter $\gamma_2$ characterizes the slope of the mean function $\eta$. We assume that the parameter $\theta=(\sigma^2,\vartheta_1, \vartheta_2)^T $ is included in every candidate model, whereas the components of the parameter $\gamma= (\gamma_1, \gamma_2)^T$ can be fixed to the corresponding components of  $\gamma_0=(0, 1)^T$, such that there are $r=4$ competing models in the   candidate set $\mathcal{S}$, that is
\begin{align} \label{logmod_narrow}
\eta_{S_1}(x,\vartheta)&= \frac{\vartheta_1 }{1+ \exp[(\vartheta_2-x)]},\\
\label{logmod_between}
\eta_{S_2}(x, \vartheta, (0,\gamma_2)^T )  &=
\frac{\vartheta_1 }{1+ \exp[(\vartheta_2-x)/\gamma_2]}, \\
 \eta_{S_3}(x, \vartheta, (\gamma_1,1)^T ) &= \gamma_1 + \frac{\vartheta_1 }{1+ \exp[(\vartheta_2-x)]}
 \label{logmod_between1}.
\end{align}
and $\eta_{S_4}$ defined by \eqref{logmod}. As the parameters $\gamma_1$ and $\vartheta_1$ appear linear in the model only the prior distributions  for $\gamma_2$ and $\vartheta_2$
have to be specified, which are chosen as independent uniform priors supported on the sets $\{3, 4, 5\}$ and $\{5/6,1,7/6\}$, respectively. The variance $\sigma^2$ is fixed as $\sigma_0^2=4.5$ and $\delta$ is chosen such that $\delta^T/\sqrt{150}=(0.015,-1/6)$. 
\\
The Bayesian optimal design for model averaging estimation  of  the AUC with equal  weights  $g_{S_i}=0.25, i=1,\ldots,4,$
has been calculated numerically and
is given by
\begin{align} \label{opt_des_log}
\xi^*_C = \left\{
\begin{array}{ccccc}
0 & 2.585 & 4.332 & 5.419 & 8 \\
0.094 & 0.258 & 0.239 & 0.204 & 0.206
\end{array}
\right\}.
\end{align}
The performance of the different  designs is again evaluated by means of a simulation study generating data
from the model
\begin{align} \label{sim_values_log}
y_{ij}^{(l)} = \gamma_1 + \frac{\vartheta_1 }{1+ \exp[(\vartheta_2-x)/\gamma_2]} + \sigma \varepsilon_{ij}^{(l)} , i=1,...,k, j=1,...,n_i,
\end{align}
where $\varepsilon_{ij}^{(l)}$  are standard normal distributed random variables  and   $n=\sum_{i=1}^k n_i=150$ observations can be taken.
We focus on the case  $\vartheta^T=(\vartheta_1,\vartheta_2) = (-1.73,4)$,   $\gamma^T= (0.015,0.833)$ and $\sigma^2=4.5$
 which corresponds to  a local misspecification, where $\theta_0^T=(4.5,-1.73,4), \ \gamma_0^T=(0,1)$ and $\delta^T/\sqrt{150}=(0.015,-1/6)$.
 Further results for other parameter choices show a similar picture and are given and discussed in Section \ref{sec63} of the appendix.
\\
 \begin{table}[t]
 \centering
 \begin{tabular}{| c || c|c|c|}
 \hline
 & \multicolumn{3}{|c|}{estimation method } \\
\cline{2-4}
design &  fixed weights &  smooth AIC & model selection  \\
 \hline\cline{2-4}
$\xi^*_C$  & \textbf{1.659} & 1.880 & 2.074 \\
 $\xi_1$ & 1.961 & 2.080 & 2.196 \\
 $\xi_2$ & 1.687 & \textbf{1.763} & \textbf{1.838} \\
  \hline
 \end{tabular}
 \caption{\it
 The    mean squared error of the model averaging estimators with weights $g_{S_i}=0.25$, $i=1, \ldots, 4$ (left column), with the smooth AIC-weights  \eqref{smooth_aic_weights} (middle column) and the estimator based on model-selection (right column).
The different rows correspond to different designs. First row: Bayesian optimal design $\xi^*_C $  for model averaging estimation of the AUC defined  in
\eqref{opt_des_log}. Middle row: uniform design  $\xi_1$  defined in \eqref{com_used_des1}. Third row:  uniform design  $\xi_2$  defined in \eqref{com_used_des2}.}
\label{simulated_mse_values_log}
 \end{table}
The mean squared error of the model averaging estimator with equal weights  $g_{S_i}=0.25$ ($i=1,\ldots,4$)  for the different designs is given in the
left column of Table \ref{simulated_mse_values_log}, while the middle and right column show the corresponding results for the model
averaging estimator with smooth AIC-weights and the estimator based on model selection, respectively.
We observe again that model averaging improves the estimation of the target AUC in all cases under consideration. For fixed weights this improvement varies between $8\%$ and
$20\%$ (depending on the design), while the improvement achieved by model averaging with smooth AIC-weights   varies between $4\%$ and
$9\%$. The model averaging estimator with fixed (equal) weights performs substantially better  than the procedure with smooth AIC-weights.
\\
In the case of fixed weights the Bayesian optimal design $\xi_C^*$
for model averaging estimation of the AUC yields a $15\%$ improvement of the the uniform design $\xi_1$ but only a $2\%$ improvement of the
design $\xi_2$. On the  other hand, if model averaging estimates with smooth AIC-weights or model selection weights are   used, the uniform design $\xi_2$ shows the best performance.
This observation can be explained by the fact that the design $\xi_C^*$ has not been constructed  for this purpose. Consequently, although this design performs very well in many cases, it cannot be guaranteed that the design $\xi_C^*$  is close to the optimal design for model averaging estimation of the AUC with smooth AIC-weights or for the estimation in a model selected by the AIC.
Nevertheless, model averaging with fixed weights and the corresponding Bayesian optimal design yields the  smallest
mean squared error in all considered scenarios.
\\
 \begin{table}[t]
 \centering
 \begin{tabular}{| c || c|c|c|}
 \hline
 & \multicolumn{3}{|c|}{estimation method } \\
\cline{2-4}
design &  fixed weights &  smooth AIC & model selection \\
 \hline\cline{2-4}
$\xi^*_D$  & \textbf{1.764} & \textbf{1.723} & \textbf{1.835} \\
 $\xi_1$ & 2.059 & 2.041 & 2.129 \\
 $\xi_2$ & 1.841 & 1.801 & 1.883 \\
  \hline
 \end{tabular}
 \caption{\it
 The    mean squared error of the model averaging estimators with weights $g_{S_1}=0.1$, $g_{S_2}=0.1$, $g_{S_3}=0.1$ and $g_{S_4}=0.7$ (left column), with the smooth AIC-weights  \eqref{smooth_aic_weights} (middle column) and the estimator based on model-selection (right column).
The different rows correspond to different designs. First row: Bayesian optimal design $\xi^*_D $  for model averaging estimation of the AUC defined  in
\eqref{opt_des_log_2}. Middle row: uniform design  $\xi_1$  defined in \eqref{com_used_des1}. Third row:  uniform design  $\xi_2$  defined in \eqref{com_used_des2}.}
\label{simulated_mse_values_log_2}
 \end{table}Next we consider a model averaging estimator with (non-uniform)  weights
$g_{S_1}=0.1 $, $g_{S_2}=0.1 $, $g_{S_3}=0.1 $, and $g_{S_4}=0.7 $  for the models   \eqref{logmod_narrow}, \eqref{logmod_between}, \eqref{logmod_between1} and \eqref{logmod}, respectively.
The corresponding Bayesian optimal design for model averaging estimation of the AUC  with these weights is  is given by
\begin{align} \label{opt_des_log_2}
\xi^*_D = \left\{
\begin{array}{ccccc}
0 & 2.418 & 4.259 & 5.777 & 8 \\
0.122 & 0.284 & 0.197 & 0.253 & 0.145
\end{array}
\right\}.
\end{align}
The   mean squared error  of the model averaging estimators for different designs
is given in the left column of  Table \ref{simulated_mse_values_log_2}, where we use the same parameters as in the previous example. The middle and right column show the simulated
mean squared error for  model averaging estimation with  smooth AIC-weights and the estimator based on model selection, respectively.
We observe a similar behaviour as described in Section \ref{sec41}:  model averaging performs better
than model selection but in this situation model averaging based on smooth AIC-weights results in a slightly
smaller mean squared error than model averaging based on fixed weights (the estimator with fixed weights yields an increase of the   mean squared error of about $2 \%$).
For all three estimators the mean squared error from  the Bayesian  optimal design
$\xi_D^*$ defined in  \eqref{opt_des_log_2} is smaller than the ones obtained from
the designs $\xi_1$ and $\xi_2$.
\\
Further simulation results using  other parameter combinations can be found in  Table \ref{tab_log_auc_150}  (model averaging estimator with equal weights)
and  Table \ref{tab_log_auc_150_2}   (model averaging estimator with non-uniform weights)   in the appendix
and show a similar picture as described in the previous paragraphs.
For example, model averaging shows a better performance than estimation in a model identified by the AIC, independently of the design under consideration.
In most cases the Bayesian optimal design for model averaging estimation  of the AUC  yields a substantial improvement  compared to the uniform designs, even when it is used for model averaging with smooth AIC-weights or for estimation after model selection (see Section \ref{sec63} for more details).

\section{Conclusions}
 \label{sec5}

In this paper we studied the problem of constructing efficient designs for parametric regression  if model averaging is used to estimate a target 
under model uncertainty. We have developed a new optimality criterion which determines a design such that the asymptotic mean squared error 
of the estimator of the target (under local deviation from the assumed model) becomes minimal by the choice of the experimental design. 
The results are illustrated by means of a simulation study  in  the problem of estimating the effective dose and  the area under the curve.
The optimal designs  yield a substantial reduction of the mean squared error of the frequentist model averaging estimate. \\
The optimal designs constructed for model averaging with fixed weights also improve  model averaging estimates with smooth 
AIC-weights and estimates in a model selected by an information type criterion. However, it remains an open and very challenging question  
for future research  to determine optimal designs for estimation methods of this type. The asymptotic distribution of these estimators is complicated 
and has to be simulated in general for each design under consideration, which is computationally very demanding. 
A further  interesting direction of future research in this context consists in the construction and investigation of adaptive designs, which proceed in
several steps, updating the information about the models and their parameters sequentially.  \\

\bigskip

{\bf Acknowledgements} This work has been supported in part by the
Collaborative Research Center ``Statistical modeling of nonlinear
dynamic processes'' (SFB 823, Teilprojekt C2, T1) of the German Research Foundation
(DFG) and  by a grant from the National Institute of General Medical Sciences of the National
Institutes of Health under Award Number R01GM107639. The content is solely the responsibility of the authors and does not necessarily
 represent the official views of the National
Institutes of Health.

\setstretch{1.15}
\setlength{\bibsep}{1pt}
\begin{small}
 \bibliographystyle{apalike}
\itemsep=0.5pt
 \bibliography{modelaveragingopt}
\end{small}

\section{Appendix} \label{sec6}
\def\theequation{A.\arabic{equation}}
\setcounter{equation}{0}

\subsection{Proof of Theorem \ref{mav_necess_cond} and Theorem \ref{bayes_mav_necess_cond}}
\label{sec61}
Theorem \ref{mav_necess_cond} is a special case of Theorem \ref{bayes_mav_necess_cond}, since the Bayesian model averaging optimality criterion reduces to the local model averaging optimality criterion with respect to the parameter $(\theta_0^T, \gamma_0^T, \delta^T)$ by choosing 
a one-point prior.  
Following the arguments in \cite{pukelsheim_optimal_2006}[Chapter 11] and assuming that integration and differentiation are interchangeable, a Bayesian optimal design $\xi^{*}$ for model averaging estimation of the parameter $\mu$ satisfies the inequality
\begin{equation}\label{nec_cond2}
-D\Phi _{\text{mav}}(\xi^*)(\xi_x-\xi^*) = -\int_{\Theta \times \Gamma } D \phimav(\xi^*,g,\delta,\theta,\gamma)(\xi_x - \xi^*) \pi(d\theta, d\gamma) \leq 0
\end{equation}
for all $x \in \mathcal{X}$, where $D \phimav(\xi^*,g,\delta,\theta_0,\gamma_0)(\xi_x - \xi^*)$ denotes the directional derivative of the function $\phimav$ evaluated in the optimal design $\xi^*$ in direction $\xi_x - \xi^*$ and $\xi_x$ denotes the Dirac measure at the point $x \in {\cal X}$. Note that in the particular case of the model averaging optimality criterion, the corresponding function  $\phimav^{\pi}(\xi)$ is not convex and therefore the necessary condition in \eqref{nec_cond2} is not sufficient. \\
We now calculate an explicit expression of the derivative using the chain rule
\begin{equation}\label{deriv_phimav}
	D \phimav(\xi^*,g,\delta,\theta,\gamma)(\xi_x - \xi^*) = 2\nu(\xi^*) D_1(\xi^*, x,\delta, \theta, \gamma) + D_2(\xi^{*}, x, \theta, \gamma),
\end{equation}
where $D_1(\xi^*, x, \delta, \theta, \gamma)$ is the directional derivative of the bias function $\nu$ defined by \eqref{mean_mav} and $D_2(\xi^{*}, x, \theta, \gamma)$ is the directional derivative of the variance function $\tau^2$ defined by \eqref{var_mav}. \\
We consider these derivatives separately starting with  $D_1(\xi^*, x, \delta, \theta, \gamma)$, for which we obtain
\begin{equation} \label{deriv_nu2}
	D_1(\xi^*, x, \delta,\theta, \gamma)= \sum_{j=1}^r g_{S_j} c^T DL_{S_j}(\xi^*,\theta,\gamma)(\xi_x - \xi^*) \delta,
\end{equation}
where $DL_{S_j}(\xi^*,\theta,\gamma)(\xi_x - \xi^*) $ denotes the derivative of the function $L_{S_j}$ defined in \eqref{L_S} and is therefore given by
\begin{equation} \label{deriv_LS}
	\begin{split}
	DL_{S_j}(\xi^*,\theta,\gamma)(\xi_x - \xi^*) = &P^T_{S_j} J_{S_j}^{-1}(\xi^*, \theta, \gamma_{S_j}) \left( P_{S_j} J(\xi_x, \theta, \gamma) \right. \\ & \left. - J_{S_j}(\xi_x, \theta, \gamma_{S_j}) J_{S_j}^{-1}(\xi^*, \theta, \gamma_{S_j})P_{S_j} J(\xi^*, \theta, \gamma) \right) \begin{pmatrix} 0 _{p\times q} \\ I_{q\times q} \end{pmatrix}.
	\end{split}
\end{equation}
Here we used that the derivative of the inverse of the information matrix, $J^{-1}_S(\xi^*)$, in direction $\xi^*-\xi_x$ is of the form
\begin{equation}\label{inv_info_deriv}
	D J_S^{-1}(\xi^*,\theta,\gamma_{S})(\xi_x - \xi^*) = J_S^{-1}(\xi^*,\theta,\gamma_{S}) - J_S^{-1}(\xi^*,\theta,\gamma_{S}) J_S(\xi_x,\theta,\gamma_{S}) J_S^{-1}(\xi^*,\theta,\gamma_{S}),
\end{equation}
for an arbitrary $S\subset\{1, \ldots, q \}$. Combining \eqref{deriv_nu2} and \eqref{deriv_LS} follows in the representation of $D_1$ given in \eqref{deriv_nu}. \\
The derivative $D_2(\xi^{*}, x, \theta, \gamma)$ is of the form
\begin{equation}\label{deriv_tau2}
	\begin{split}
	D_2(\xi^{*}, x, \theta, \gamma)= \sum_{i,j=1}^{r}  g_{S_i} g_{S_j} & \left( 2D h^T_{S_i}(\xi^*)(\xi_x-\xi^*) J(\xi^*,\theta,\gamma) h_{S_j}(\xi^*) \right. \\
	& \left. + h_{S_i}^T(\xi^*)  \{J(\xi^*, \theta, \gamma)+ J(\xi_x, \theta, \gamma)\}  h_{S_j}(\xi^*) \right),
	\end{split}
\end{equation}
where $D h_{S}(\xi^*)(\xi_x-\xi^*)$ denotes the derivative of $h_{S}$ defined by \eqref{hs} for an aribitrary subset $S\subset \{1, \ldots, q\}$. Using \eqref{inv_info_deriv} $D h_{S}(\xi^*)(\xi_x-\xi^*)$ is given by
\begin{equation}\label{hs_tilde2}
	D h_{S}(\xi^*)(\xi_x-\xi^*) =  h_{S}(\xi^*) - \tilde{h}_{S}(\xi^*, \xi_x)
\end{equation}
where $\tilde{h}_{S}$ is defined by
$$ \tilde{h}_{S}(\xi^*, \xi_x) =  P^T_S J^{-1}_{S}(\xi^*, \theta, \gamma_{S}) J_{S}(\xi_x, \theta, \gamma_{S}) J_{S}^{-1}(\xi^*, \theta, \gamma_{S})  c_{S}.$$
Combining \eqref{deriv_tau2} and \eqref{hs_tilde2} results in the representation of $D_2$ given in \eqref{deriv_tau}.
Finally, the necessary condition in \eqref{nec_cond} follows by the combination of \eqref{nec_cond2} and \eqref{deriv_phimav}.
\medskip 

To prove that there holds equality in \eqref{nec_cond_bayes} for all support points $x$ of the design $\xi^*$, assume that there exists at least one support point $x_0$ of the design $\xi^*$, such that the inequality in \eqref{nec_cond_bayes} is strict. Then, we have
\begin{align*}
	\int_{\mathcal{X}} \int_{\Theta \times \Gamma } \left( -2\nu(\xi^*, \delta, \theta, \gamma) D_1(\xi^*, x,\delta,\theta, \gamma)   -   D_2(\xi^*, x,\theta, \gamma)\right) \pi(d\theta, d\gamma) \xi^*(dx) < 0. 
\end{align*}
On the other hand, since $\int_{\mathcal{X}} J(\xi_x,\theta,\gamma) \xi^*(dx) = J(\xi^*,\theta,\gamma)$ and $\int_{\mathcal{X}} \tilde{h}_{S}(\xi^*, \xi_x) \xi^*(dx) = h_{S}(\xi^*)$, we have
\begin{align*}
	\int_{\mathcal{X}} D_1(\xi^*, x,\delta,\theta, \gamma) \xi^*(dx) = 0 \text{ and } \int_{\mathcal{X}} D_2(\xi^*, x,\theta, \gamma) \xi^*(dx) = 0,
\end{align*}
such that 
\begin{align*}
	\int_{\mathcal{X}} &\int_{\Theta \times \Gamma } -2\nu(\xi^*, \delta, \theta, \gamma) D_1(\xi^*, x,\delta,\theta, \gamma)   -   D_2(\xi^*, x,\theta, \gamma) \pi(d\theta, d\gamma) \xi^*(dx) \\ 
	&=  
	\int_{\Theta \times \Gamma } \left\{ -2\nu(\xi^*, \delta, \theta, \gamma) \int_{\mathcal{X}} D_1(\xi^*, x,\delta,\theta, \gamma) \xi^*(dx)  - \int_{\mathcal{X}}  D_2(\xi^*, x,\theta, \gamma) \xi^*(dx) \right\} \pi(d\theta, d\gamma) = 0,
\end{align*}
which is a contradiction. Consequently, equality in \eqref{nec_cond_bayes} must hold whenever $x$ is a support point of the design $\xi^*$.

\subsection{Additional simulation results}\label{sec6add}

\subsubsection{Estimation of  the $\text{ED}_{0.6}$}\label{sec62}

\begin{table}[!t]
\footnotesize
\begin{center}
 \begin{tabular}{| c | c || c|c|c|}
 \hline
 Parameter & design  & \multicolumn{3}{|c|}{estimation method} \\
\cline{3-5}
 $(\vartheta,\gamma)$ & &  fixed weights &  smooth AIC & model selection  \\
 \hline\cline{3-5}
& $\xi_A^*$  & \textbf{0.818 }& \textbf{1.065} & \textbf{1.180} \\
(1.81,0.79,0,1) & $\xi_1$ & 1.339 & 1.526 & 1.660 \\
 & $\xi_2$ & 1.207 & 1.549 & 1.791 \\
  \hline
 & $\xi_A^*$  & \textbf{0.718} & \textbf{0.957} & \textbf{1.059} \\
(1.81,0.79,0.1,1) & $\xi_1$ & 1.238 & 1.413 & 1.535 \\
 & $\xi_2$ & 1.045 & 1.406 & 1.695 \\
 \hline
 & $\xi_A^*$  & \textbf{0.394} & \textbf{0.533} & \textbf{0.639} \\
(1.81,0.79,0,2) & $\xi_1$ & 0.788 & 0.823 & 0.915 \\
 & $\xi_2$ & 0.659 & 0.852 & 0.975 \\
 \hline
& $\xi_A^*$  & \textbf{0.355} & \textbf{0.508} & \textbf{0.596} \\
(1.81,0.79,0.1,2) & $\xi_1$ & 0.810 & 0.913 & 1.017 \\
 & $\xi_2$ & 0.637 & 0.846 & 0.994 \\
 \hline
 & $\xi_A^*$  & \textbf{0.732} & \textbf{0.953} & \textbf{1.103} \\
(1.81,1.79,0,2) & $\xi_1$ & 1.374 & 1.570 & 1.767 \\
 & $\xi_2$ & 1.119 & 1.437 & 1.660 \\
  \hline
 & $\xi_A^*$  & \textbf{0.777} & \textbf{1.121} & 1.453 \\
(1.81,1.79,0.1,2) & $\xi_1$ & 1.166 & 1.384 & 1.532 \\
 & $\xi_2$ &  0.985 & 1.222 & \textbf{1.415} \\
   \hline
 & $\xi_A^*$  & \textbf{0.449} & \textbf{0.513} & \textbf{0.623} \\
(1.81,1.79,0,3) & $\xi_1$ & 0.988 & 1.144 & 1.250 \\
 & $\xi_2$ & 0.762 & 0.908 & 1.049 \\
    \hline
 & $\xi_A^*$  & \textbf{0.464} & \textbf{0.598} & \textbf{0.713} \\
(1.81,1.79,0.1,3) & $\xi_1$ & 0.932 & 1.182 & 1.314 \\
 & $\xi_2$ & 0.724 & 0.892 & 1.061 \\
 \hline
 \end{tabular}
 \end{center}
 \caption{\it
 The    mean squared error of the model averaging estimators  of the $\text{ED}_{0.6}$ with weights $g_{S_i}=0.25, i=1,\ldots,4$ (left column), with the smooth AIC-weights  \eqref{smooth_aic_weights} (middle column) and the estimator based on model-selection (right column).
The different rows correspond to different parameter combinations. Within each parameter combination the different rows correspond to different designs. First row: Bayesian optimal design $\xi^*_A $  for model averaging estimation of the $\text{ED}_{0.6}$ defined in \eqref{opt_des_emax}. Middle row: uniform design  $\xi_1$  defined in \eqref{com_used_des1}. Third row:  uniform design  $\xi_2$  defined in \eqref{com_used_des2}.}
\label{tab_emax_edp_150}
 \end{table}

\begin{table}[!t]
\footnotesize
 \begin{center}
 \begin{tabular}{| c | c || c|c|c|}
 \hline
 Parameter &design  & \multicolumn{3}{|c|}{estimation method} \\
\cline{3-5}
 $(\vartheta,\gamma)$ &  &  fixed weights &  smooth AIC& model selection  \\
 \hline\cline{3-5}
& $\xi_B^*$  & \textbf{0.864} & \textbf{0.849} & \textbf{1.012} \\
(1.81,0.79,0,1) & $\xi_1$ & 1.504 & 1.498 & 1.605 \\
 & $\xi_2$ & 1.382 & 1.450 & 1.631 \\
  \hline
 & $\xi_B^*$  & \textbf{0.914} & \textbf{0.937} & \textbf{1.112} \\
(1.81,0.79,0.1,1) & $\xi_1$ & 1.493 & 1.497 & 1.613 \\
 & $\xi_2$ & 1.306 & 1.310 & 1.491 \\
 \hline
 & $\xi_B^*$  & \textbf{0.540} & \textbf{0.536} & \textbf{0.600} \\
(1.81,0.79,0,2) & $\xi_1$ & 0.967 & 0.967 & 1.048 \\
 & $\xi_2$ & 0.834 & 0.861 & 1.004 \\
 \hline
& $\xi_B^*$  & \textbf{0.476} & \textbf{0.502} & \textbf{0.582} \\
(1.81,0.79,0.1,2) & $\xi_1$ & 0.915 & 0.900 & 1.014 \\
 & $\xi_2$ & 0.869 & 0.949 & 1.067 \\
 \hline
 & $\xi_B^*$  & \textbf{0.904} & \textbf{0.873} & \textbf{1.038} \\
(1.81,1.79,0,2) & $\xi_1$ &  1.292 & 1.329 & 1.506 \\
 & $\xi_2$ & 1.362 & 1.338 & 1.611 \\
  \hline
 & $\xi_B^*$  & \textbf{0.875} & \textbf{0.931} &\textbf{ 1.091} \\
(1.81,1.79,0.1,2) & $\xi_1$ & 1.382 & 1.410 & 1.573 \\
 & $\xi_2$ & 1.350 & 1.368 & 1.599 \\
   \hline
 & $\xi_B^*$  &\textbf{ 0.516} & \textbf{0.532} & \textbf{0.619} \\
(1.81,1.79,0,3) & $\xi_1$ & 1.129 & 1.144 & 1.251 \\
 & $\xi_2$ & 0.836 & 0.813 & 0.927 \\
    \hline
 & $\xi_B^*$  & \textbf{0.578} & \textbf{0.560} & \textbf{0.615} \\
(1.81,1.79,0.1,3) & $\xi_1$ & 1.130 & 1.171 & 1.304 \\
 & $\xi_2$ & 0.800 & 0.851 & 1.023 \\
 \hline
 \end{tabular}
  \end{center}
 \caption{\it
 The    mean squared error of the model averaging estimators  of the $\text{ED}_{0.6}$
 with weights $g_{S_1}=0.1, g_{S_2}=0.1,g_{S_3}=0.3$ and $g_{S_4}=0.5$ (left column), with the smooth AIC-weights  \eqref{smooth_aic_weights} (middle column) and the estimator based on model-selection (right column).
The different rows correspond to different parameter combinations. Within each parameter combination the different rows correspond to different designs. First row: Bayesian optimal design $\xi^*_B $  for model averaging estimation of the $\text{ED}_{0.6}$ defined in \eqref{opt_des_emax_2}. Middle row: uniform design  $\xi_1$  defined in \eqref{com_used_des1}. Third row:  uniform design  $\xi_2$  defined in \eqref{com_used_des2}.}
\label{tab_emax_edp_150_2}
 \end{table}

In this section we present further simulation results for the estimation of the $\text{ED}_{0.6}$ in the sigmoid Emax model. Data is generated from the model
\eqref{sim_values} where $n=150$ observations  are taken according to the designs  $\xi_A^*$,  $\xi_B^*$, $\xi_1$ and $\xi_2$ defined in Section \ref{sec41}.
Different parameters  $(\vartheta,\gamma)$ are considered to demonstrate that the results in Section \ref{sec41} are representative.
The simulated mean squared error for the model averaging estimates of the  $\text{ED}_{0.6}$ can be found in Table
\ref{tab_emax_edp_150} (uniform weights $g_{S_i}=0.25, i=1,\ldots,4$) and Table \ref{tab_emax_edp_150_2} (non-uniform weights $g_{S_1}=0.1, g_{S_2}=0.1,g_{S_3}=0.3$ and $g_{S_4}=0.5$). In the left and middle column we display the results
for  the model averaging estimator of the $\text{ED}_{0.6}$ with fixed weights  and with smooth AIC-weights, respectively,
while the right column shows the results for estimation of the $\text{ED}_{0.6}$ in the model selected via AIC.
 \\
We observe from Table \ref{tab_emax_edp_150} that model averaging estimation
 always yields a smaller   mean squared error than estimation after model selection via AIC. 
Model averaging estimation with  fixed weights results in a
reduction of the mean squared error by $14\%$-$47\%$ whereas  smooth AIC-weights reduce the mean squared error by $7\%$-$23\%$. Moreover, model averaging with fixed weights shows a better performance than
model averaging  with data driven smooth AIC-weights. 
Table \ref{tab_emax_edp_150_2} shows similar results for model averaging estimation with non-uniform weights, but  the difference between model averaging estimation
with  fixed weights and data driven weights is less substantial. Moreover,  there are also a few parameter combinations where using
non-uniform fixed weights yields a slight  increase of the  mean squared error  (about $1-3\%$) compared to
 smooth AIC-weights.
 \\
Next we compare  the optimal designs $\xi_A^*$ and $\xi_B^*$ with the uniform designs $\xi_1$ and $\xi_2$
which yield  a reduction of the   mean squared error of the  model averaging estimator of the $\text{ED}_{0.6}$ with fixed weights by $21\%$-$56\%$ and by $28\%$-$54\%$, respectively.
For model averaging estimation with smooth AIC-weights the optimal designs $\xi_A^*$ and $\xi_B^*$ reduce the mean squared error by $8\%$-$55\%$ and $28\%$-$53\%$, respectively.  Finally, for  estimation  of the $\text{ED}_{0.6}$ in the model identified by the AIC the optimal designs  reduce the mean squared error
in almost all considered cases.

\subsubsection{Estimation of the $\text{AUC}$}\label{sec63}
\begin{table}[!t]
\footnotesize
 \begin{center}
 \begin{tabular}{| c | c || c|c|c|}
 \hline
 Parameter & design  & \multicolumn{3}{|c|}{estimation method} \\
\cline{3-5}
 $(\vartheta,\gamma)$ & &  fixed weights &  smooth AIC & model selection \\
 \hline\cline{3-5}
& $\xi^*$  & \textbf{1.559} & \textbf{1.741} & \textbf{1.871} \\
(-1.73,4,0,1) & $\xi_1$ & 1.886 & 1.963 & 2.030 \\
 & $\xi_2$ & 1.880 & 1.959 & 2.042 \\
  \hline
 & $\xi^*$  & \textbf{1.503} & \textbf{1.658} & \textbf{1.802} \\
(-1.73,4,0.015,1) & $\xi_1$ & 2.060 & 2.140 & 2.222 \\
 & $\xi_2$ & 1.831 & 1.917 & 1.981 \\
 \hline
 & $\xi^*$  & \textbf{1.630} & 1.825 & 1.986 \\
(-1.73,4,0,0.833) & $\xi_1$ & 2.042 & 2.139 & 2.230 \\
 & $\xi_2$ & 1.681 & \textbf{1.811} & \textbf{1.883} \\
 \hline
& $\xi^*$  & \textbf{1.659} & 1.880 & 2.074 \\
(-1.73,4,0.015,0.833) & $\xi_1$ & 1.961 & 2.080 & 2.196 \\
 & $\xi_2$ & 1.687 & \textbf{1.763} & \textbf{1.838} \\
 \hline
 & $\xi^*$  & \textbf{1.442} & \textbf{1.637} & \textbf{1.762} \\
(-1.73,5,0,0.833) & $\xi_1$ & 1.671 & 1.815 & 1.925 \\
 & $\xi_2$ & 1.659 & 1.846 & 1.996 \\
  \hline
 & $\xi^*$  & \textbf{1.517} & 1.773 & 1.953 \\
(-1.73,5,0.015,0.833) & $\xi_1$ & 1.690 & 1.820 & 1.924  \\
 & $\xi_2$ & 1.629 & \textbf{1.764} & \textbf{1.884} \\
   \hline
 & $\xi^*$  & \textbf{1.389} & \textbf{1.688} & 1.873 \\
(-1.73,5,0,0.667) & $\xi_1$ & 1.672 & 1.823 & 1.955 \\
 & $\xi_2$ & 1.511 & 1.691 & \textbf{1.807} \\
    \hline
 & $\xi^*$  & \textbf{1.421} & \textbf{1.687} & \textbf{1.839} \\
(-1.73,5,0.015,0.667) & $\xi_1$ & 1.649 & 1.870 & 2.040 \\
 & $\xi_2$ & 1.626 & 1.792 & 1.907\\
 \hline
 \end{tabular}
  \end{center}
 \caption{\it
 The    mean squared error of the model averaging estimators of the AUC with weights $g_{S_i}=0.25, i=1,\ldots,4$ (left column), with the smooth AIC-weights  \eqref{smooth_aic_weights} (middle column) and the estimator based on model-selection (right column).
The different rows correspond to different parameter combinations. Within each parameter combination the different rows correspond to different designs. First row: Bayesian optimal design $\xi^*_C $  for model averaging estimation of the AUC  defined  in
\eqref{opt_des_log}. Middle row: uniform design  $\xi_1$  defined in \eqref{com_used_des1}. Third row:  uniform design  $\xi_2$  defined in \eqref{com_used_des2}.}
\label{tab_log_auc_150}
 \end{table}

\begin{table}[!t]
\footnotesize
 \begin{center}
 \begin{tabular}{| c | c || c|c|c|}
 \hline
 Parameter & design  & \multicolumn{3}{|c|}{estimation method} \\
\cline{3-5}
 $(\vartheta,\gamma)$ & &  fixed weights &  smooth AIC & model selection \\
 \hline\cline{3-5}
& $\xi^*$  & \textbf{1.913} & \textbf{1.851} & \textbf{1.956} \\
(-1.73,4,0,1) & $\xi_1$ & 2.159 & 2.128 & 2.213 \\
 & $\xi_2$ & 1.942 & 1.918 & 1.989 \\
  \hline
 & $\xi^*$  & \textbf{1.890} & \textbf{1.843} & \textbf{1.951} \\
(-1.73,4,0.015,1) & $\xi_1$ & 2.042 & 2.018 & 2.106 \\
 & $\xi_2$ & 1.935 & 1.912 & 1.959 \\
 \hline
 & $\xi^*$  & \textbf{1.662} & \textbf{1.604} & \textbf{1.702} \\
(-1.73,4,0,0.833) & $\xi_1$ & 1.964 & 1.934 & 2.025 \\
 & $\xi_2$ & 1.832 & 1.807 & 1.875 \\
 \hline
& $\xi^*$  & \textbf{1.764} & \textbf{1.723} & \textbf{1.835} \\
(-1.73,4,0.015,0.833) & $\xi_1$ & 2.059 & 2.041 & 2.129 \\
 & $\xi_2$ & 1.841 & 1.801 & 1.883 \\
 \hline
 & $\xi^*$  & 1.863 & \textbf{1.771} & \textbf{1.886} \\
(-1.73,5,0,0.833) & $\xi_1$ & 1.881 & 1.818 & 1.930 \\
 & $\xi_2$ & \textbf{1.842} & 1.813 & 1.942 \\
  \hline
 & $\xi^*$  & \textbf{1.689} & \textbf{1.617} &\textbf{ 1.761} \\
(-1.73,5,0.015,0.833) & $\xi_1$ & 2.006 & 1.944 & 2.083  \\
 & $\xi_2$ & 1.700 & 1.670 & 1.815 \\
   \hline
 & $\xi^*$  & \textbf{1.671} & \textbf{1.590} & \textbf{1.716} \\
(-1.73,5,0,0.667) & $\xi_1$ & 1.833 & 1.769 & 1.925 \\
 & $\xi_2$ & 1.818 & 1.768 & 1.920 \\
    \hline
 & $\xi^*$  & 1.745 & 1.665 & 1.816 \\
(-1.73,5,0.015,0.667) & $\xi_1$ & 1.896 & 1.824 & 1.957 \\
 & $\xi_2$ & \textbf{ 1.649} & \textbf{1.626} & \textbf{1.779}\\
 \hline
 \end{tabular}
  \end{center}
 \caption{\it
 The    mean squared error of the model averaging estimators of the AUC with weights $g_{S_1}=0.1, g_{S_2}=0.1, g_{S_3}=0.1$ and $g_{S_4}=0.7$ (left column), with the smooth AIC-weights  \eqref{smooth_aic_weights} (middle column) and the estimator based on model-selection (right column).
The different rows correspond to different parameter combinations. Within each parameter combination the different rows correspond to different designs. First row: Bayesian optimal design $\xi^*_D $  for model averaging estimation of the AUC defined  in
\eqref{opt_des_log_2}. Middle row: uniform design  $\xi_1$  defined in \eqref{com_used_des1}. Third row:  uniform design  $\xi_2$  defined in \eqref{com_used_des2}.}
\label{tab_log_auc_150_2}
 \end{table}
 
In this section we present further simulation results for model averaging estimation of the AUC in the logistic model. We generate data from the model
\eqref{sim_values_log} where $n=150$ observations  are taken according to the designs   $\xi_1$ and $\xi_2$, $\xi_C^*$,  $\xi_D^*$ defined in Section \ref{sec42}.  To validate the findings in Section \ref{sec42} for other choices of the parameter we consider further scenarios for the
 parameter  $(\vartheta,\gamma)$ and
 simulate the mean squared error of the model averaging estimators of the AUC. The results  can be found in Table
\ref{tab_log_auc_150} (uniform weights $g_{S_i}=0.25, i=1,\ldots,4$) and Table \ref{tab_log_auc_150_2} (non-uniform weights $g_{S_1}=0.1, g_{S_2}=0.1,g_{S_3}=0.1$ and $g_{S_4}=0.7$). In the  left column of these tables
we display the  results of the model averaging estimator of the AUC with fixed weights, while the middle and right column
show the mean squared error of the model averaging estimator  with smooth AIC-weights and the estimator
based on  model selection, respectively.  \\
As in Section \ref{sec42}  we observe that the   mean squared error of model averaging estimators is always smaller than the   mean squared error of estimators after model selection via AIC (improvement: $7\%$-$26\%$ with uniform weights, $1\%$-$7\%$ with non-uniform weights and $2\%$-$10\%$ with smooth AIC-weights). Model averaging estimation of the AUC with uniform weights yields  a reduction of the mean squared error by $4\%$-$18\%$ (depending on the design and parameters) compared to model averaging estimation  with smooth AIC-weights [see Table \ref{tab_log_auc_150}]. On the other hand
 non-uniform weights yield a slight  increase of the mean squared error  [see Table \ref{tab_log_auc_150_2}].
 \\
We observe from Table \ref{tab_log_auc_150} that the Bayesian optimal designs  improve the uniform designs  for model averaging estimation of the AUC with  uniform weights  in all scenarios under consideration  (improvement:  $2\%$-$27\%$). For the estimator with
 non-uniform weights the improvement varies between $1\%$-$16\%$, although there are a few parameter combinations with no improvement [see Table \ref{tab_log_auc_150_2}].
 For model averaging  with data driven weights the optimal design  $\xi_C^*$  (constructed for fixed weights)  improves
the uniform design  $\xi_2$ in roughly half of the scenarios under consideration and the  the Bayesian optimal design $\xi_D^*$
determined for non-uniform weights performs better than $\xi_1$ and $\xi_2$ in most cases.

\end{document}